\documentclass{bioinfo2}
\copyrightyear{2012}
\pubyear{2012}

\usepackage[ruled,vlined]{algorithm2e}

\usepackage{natbib}
\begin{document}
\firstpage{1}

\title[Variant calling from de novo assembly]{Exploring single-sample SNP and INDEL calling with whole-genome de novo assembly}

\author[Li]{Heng Li$^{1,}$\footnote{to whom correspondence should be addressed}}

\address{$^1$Broad Institute, 7 Cambridge Center, Cambridge, MA 02142, USA}

\history{Received on XXXXX; revised on XXXXX; accepted on XXXXX}
\editor{Associate Editor: XXXXXXX}
\maketitle

\begin{abstract}

\section{Motivation:}
Eugene Myers in his string graph paper~\citep{Myers:2005bh} suggested that in a
string graph or equivalently a unitig graph, any path spells a valid assembly.
As a string/unitig graph also encodes every valid assembly of reads, such a
graph, provided that it can be constructed correctly, is in fact a lossless
representation of reads. In principle, every analysis based on whole-genome
shotgun sequencing (WGS) data, such as SNP and insertion/deletion (INDEL)
calling, can also be achieved with unitigs.

\section{Results:}
To explore the feasibility of using de novo assembly in the context of
resequencing, we developed a de novo assembler, \emph{fermi}, that assembles
Illumina short reads into unitigs while preserving most of information of the
input reads. SNPs and INDELs can be called by mapping the unitigs against a
reference genome. By applying the method on 35-fold human resequencing data,
we showed that in comparison to the standard pipeline, our approach yields
similar accuracy for SNP calling and better results for INDEL calling. It has
higher sensitivity than other de novo assembly based methods for variant
calling. Our work suggests that variant calling with de novo assembly can be a
beneficial complement to the standard variant calling pipeline for whole-genome
resequencing. In the methodological aspects, we proposed FMD-index for
forward-backward extension of DNA sequences, a fast algorithm for finding
all super-maximal exact matches and one-pass construction of unitigs
from an FMD-index.

\section{Availability:} \href{http://github.com/lh3/fermi}{http://github.com/lh3/fermi}
\section{Contact:} hengli@broadinstitute.org
\end{abstract}

\vspace*{-1em}

\section{INTRODUCTION}

The rapidly decreasing sequencing cost has enabled whole-genome shotgun (WGS)
resequencing at an affordable price. Many software packages have been developed
to call variants, including SNPs, short insertions and deletions (INDELs) and
structural variations (SVs), from WGS data. At present, the standard approach to
variant calling is to map raw sequence reads against a reference genome and then
to detect differences from the reference. It is well established and has been
proved to work from a single sample to thousands of
samples~\citep{1000-Genomes-Project-Consortium:2010qc}. Nonetheless, a
fundamental flaw in this mapping based approach is that mapping algorithms
ignore the correlation between sequence reads. They are unable to take full
advantage of data and may produce inconsistent outputs which complicate variant
calling.  This flaw has gradually attracted the attention of various research
groups who subsequently proposed several methods to alleviate the effect,
including post alignment
filtering~\citep{Li:2008zr,Ossowski:2008if,Krawitz:2010zr}, iterative
mapping~\citep{Manske:2009ve}, read
realignment~\citep{Albers:2010ud,Homer:2010kx,Li:2011kx,Depristo:2011vn} and
local assembly~\citep{Carnevali:2011fk}.  However, because these methods still
rely on the initial mapping, it is difficult for them to identify and recover
mismapped or unmapped reads due to high sequence divergence, long insertions,
SVs, copy number changes or misassemblies of the reference genome. They have
not solved the problem from the root.

Another distinct approach to variant calling that fundamentally avoids the flaw of the
mapping based approach is to assemble sequence reads into contigs and to
discover variants via assembly-to-assemby alignment. It was probably more
widely used in the era of capillary sequencing. The assembly based method
became less used since 2008 due to the great difficulties in assembling 25bp
reads, but with longer paired-end reads and improved methodology,
de novo assembly is reborn as the preferred choice for variant discovery
between small genomes.

For variant discovery between human genomes, however, the assembly based
approach has not attracted much attention. Assembling a human genome is far
more challenging than assembling a bacterial genome, firstly due to the sheer
size of the genome, secondly to the rich repeats and thirdly due to the
diploidy of the human genome. Many heuristics effective for assembling small
genomes are not directly applicable to the human genome assembly. As a result,
only a few de novo assemblers have been applied on human short-read data. Among
them, ABySS~\citep{Simpson:2009ys}, SOAPdenovo~\citep{Li:2010vn} and
SGA~\citep{Simpson:2011ly}, as of now, do not explicitly output heterozygotes.
Although in theory it is possible to recover heterozygotes from their
intermediate output, it may be difficult in practice as the assemblers may not
distinguish heterozygotes from sequencing errors. Cortex~\citep{Iqbal:2012ys}
is specifically designed for retaining heterozygous variants in an assembly,
but it may be missing heterozygotes. ALLPATHS-LG~\citep{Gnerre:2011ys}
also paid particular attention to keep heterozygotes, but it still has
relatively a low sensitivity. In addition, ALLPATHS-LG only works with
reads from libraries with distinct insert size distributions and prefers read
pairs with mean insert size below three times of the read length, while many
resequencing projects do not meet these requirements and thus ALLPATHS-LG may
not be applied or work to the best performance. Even if we also include de novo
assemblers developed for capillary sequence reads, the version of the Celera
assembler used for assembling the HuRef genome~\citep{Levy:2007uq} is the only
one that retains heterozygotes while capable of assembling a mammalian genome.
At last, one may think to map sequence reads back to the assembled contigs to
recover heterozygous events, but this
procedure will be affected by the same flaw of read mapping. To the best of our
knowledge, no existing de novo assemblers are able to achieve the sensitivity
of the standard mapping based approach for a diploid mammalian genome.

In this article, we will show the first time that the assembly based variant
calling can achieve a SNP accuracy close to the standard mapping approach
and have particular strength in INDEL calling, confirming previous
studies~\citep{Iqbal:2012ys}. In addition, the de novo assembly
algorithm, \emph{fermi}, developed for this practice is also a capable
assembler for human assembly.

\vspace*{-1em}

\begin{methods}
\section{METHODS}

The methods section is organized as follows. We first review the history
of de novo assembly in the theoretical aspects, which leads to the rationale
behind fermi: to use unitigs as a lossless representation of reads. We then
summarize the notations used in the article and introduce bidirectional
FM-index for DNA sequences. We will present several algorithms for assembling
using the bidirectional FM-index. The key algorithm is based on previous
works~\citep{Simpson:2010uq}, but we need to adapt it to our new index. We
also remove the recursion in the original algorithm. Finally we will discuss
practical concerns in implementation.

\subsection{Theoretical background}
\subsubsection{A history of the overlap-layout-consensus paradigm}
Computer assisted sequence assembly can be dated back to the late
1970s~\citep{Staden:1979dq,Gingeras:1979cr}. 
In 1984, \citeauthor{Peltola:1984qf} first formulated the DNA assembling
problem as finding the shortest string (the assembly) such that each sequence
read can be mapped to the assembly within a required error rate.
To solve the problem, they proposed a three-step procedure, which is
essentially the overlap-layout-consensus (OLC) approach.

\citet{Myers:1995nx} pointed out that reducing DNA assembly to a shortest
string problem is flawed in the presence of repeat. He (see also
\citealt{DBLP:journals/algorithmica/KececiogluM95}) further proposed the
concept of \emph{overlap graph}, where a vertex corresponds to a read and a
bidirectional edge to an overlap. Naively, the DNA assembling problem can be
cast as finding a path in the overlap graph such that each vertex/read is
visited exactly once (though edge/overlap caused by repeats are not required to
be traversed), equivalent to a Hamilton path problem which is known to be
NP-complete. This has led many to believe that the OLC approach is
theoretically crippled.

However, this is a misbelief. Although the assembly problem can be reduced to a
Hamilton path problem, it can be reduced to other problems as well and in
practice almost no assemblers try to solve a Hamilton path problem.  We note
that a fundamental difference between a generic graph and an overlap graph is
that the latter can be transitively reduced while retaining the read
relationship. More formally, if $v_1\to v_2$, $v_2\to v_3$ and $v_1\to v_3$ are
all present, edge $v_1\to v_3$ is said to be \emph{reducible}. When we removed
all the contained reads and reducible edges, a procedure called
\emph{transitive reduction}, the resulting graph is still a loyal
representation of the overlap graph~\citep{Myers:1995nx}, but the path
corresponding to the assembly is not a Hamilton path any more because reads
from repetitive regions need to be traversed multiple times.

In a transitively reduced graph, if there exists $v_1\to v_2$ with the
out-degree of $v_1$ and in-degree of $v_2$ both equal to 1, we are able to
merge $v_1$ and $v_2$ into one vertex without altering the topology of the
graph. After we performed all possible merges, we get a \emph{unitig graph} in
which each vertex corresponds to a \emph{unitig}, representing a maximal linear
sequence that can be resolved by reads. Multiple copies of a repeat may be
collapsed to a single unitig. The concept of unitig helps to greatly simplify
an assembly graph. It has played a central role in the Celera
assembler~\citep{Myers:2000kl}.

On the other hand, introducing unitigs has not theoretically solved the
sequence assembly to the end. \citet{Myers:2005bh}, based on the suggestion
by~\citet{Pevzner:2001vn}, proposed to compute a traversal count for each edge
and to remove false overlap edges by solving a minimum cost network flow
problem. Finding the optimal assembly in the resultant graph can be reduced to
a Eulerian path problem, which has a linear time solution. Myers originally
applied this procedure to \emph{string graph}, an equivalence to unitig graph.
\citet{Medvedev:2009ve} pointed out that determining the traversal count can
also be resolved directly in the bidirectional unitig graph using a bidirected
network flow-based algorithm.

Once we get a transitively reduced graph, the subsequent steps can be achieved
in roughly linear time most of time -- the worst case almost never happens
globally in practice. However, deriving an overlap graph takes $O(N^2)$ time,
where $N$ is the number of reads, and transitive reduction takes at least
$O(E)$ time, where $E$ is the number of edges which is usually much larger than
$N$. This still makes an OLC based approach less favorable in short-read
assembly where $N$ can be of the order of $10^9$.

A breakthrough achieved by~\citet{Simpson:2010uq} finally solved
this last remaining problem at least when we only consider exact overlaps.
These authors developed an $O(N)$ algorithm to find all the irreducible edges,
effectively replacing the overlapping and transitive reduction phases.

In summary, in the OLC paradigm, the sequence assembly problem can be
practically solved in a time roughly linear in the total length of reads.

\subsubsection{De Bruijn graph and read coherence}
De Bruijn graph is an alternative graph representation of sequence
reads~\citep{Idury:1995oq}.  It can be trivially constructed with a simple
linear-time algorithm. This makes the de Bruijn graph approach very attractive
for assembling many short reads.

However, de Bruijn is `lossy'. From a theoretical point view, a de Bruijn graph
is equivalent to an overlap graph built by splitting a long read into overlap
$k$-mers and requiring ($k$-1)-mer exact overlaps between non-redundant
$k$-mers.  Such a graph does not have transitive edges. Because long reads all
effectively work as $k$-bp reads in a de Bruijn graph, long-range information
is lost. As a result, a path in the graph may be invalidated by reads. In
contrast, in a unitig graph or equivalently a string graph each path models a
valid assembly from input reads. \citet{Myers:2005bh} called this property of
path consistency as \emph{read coherence}.

Losing long-range information in reads, a de Bruijn graph by itself has
reduced power to resolve short repeats. This flaw is usually amended by mapping
reads back to the graph and bisecting repeats shorter than the reads, a procedure
some called as \emph{read threading}. Many de Buijn graph based assemblers
essentially take this
strategy~\citep{Pevzner:2001vn,Chaisson:2009fk,Zerbino:2009ly,Li:2010vn},
though they may use different terminologies.

With read threading, it is theoretically possible to transform a de Bruijn graph
to a coherent graph, but in practice, threading is not straightforward and
may be inefficient given complex repeat structures. For a coherent representation
of reads, a unitig graph is simpler to construct.

\subsubsection{Concluding remark}
We noted that we only focused on the theoretical aspects of de novo assembly.
In practice, many assemblers derived the final assembly by applying heuristics
on the simplified graph instead of solving a network flow problem or a Eulerian
problem. Furthermore, correcting errors, utilizing read pairs and controlling memory
usage all pose challenges to large-scale de novo assembly. Many practical
problems are not solved perfectly. De novo assembly is still a field under
active development.

\subsection{Rationale}
Being coherent, a perfectly constructed unitig graph annotated with per-unitig
read counts in fact encapsulates all the information of reads and encodes no
information invalidated by reads. In this sense, any unitig based analysis
has an equivalent read based analysis, and vice versa. This article just
uses this property to explore the applications for which we usually rely on
reads.

\subsection{Strings and FM-index}

\begin{table}[tb]\label{tab:notation}
\processtable{Notations}
{\begin{tabular}{lp{7cm}}
\toprule
Symbol & Description \\
\midrule
$T$ & String: $T=a_0a_1\ldots a_{n-1}$ with $a_{n-1}=\$$\\
$|T|$ & Length of $T$ including sentinels: $|T|=n$\\
$T[i]$ & The $i$-th symbol in string $T$: $T[i]=a_i$\\
$T[i,j]$ & Substring: $T[i,j]=a_i\ldots a_j$\\
$T_i$ & Suffix: $T_i=T[i,n-1]$\\
$S$ & Suffix array; $S(i)$ is the position of the $i$-th smallest suffix\\
$B$ & BWT: $B[i]=T[S(i)-1]$ if $S(i)>0$ or $B[i]=\$$ otherwise\\
$C(a)$ & Accumul. count array: $C(a)=|\{0\le i\le n-1:T[i]<a\}|$ \\
$O(a,i)$ & Occurrence array: $O(a,i)=|\{0\le j\le i:B[j]=a\}|$\\
$P\circ W$ & String concatenation of string $P$ and $W$\\
$Pa$ & String concatenation of string $P$ and symbol $a$: $Pa=P\circ a$\\
$\overline{P}$ & Watson-Crick reverse complement of DNA string $P$\\
\botrule
\end{tabular}}{}
\end{table}

\subsubsection{Strings with multiple sentinels}

Let $\Sigma=\{\mbox{\tt \$},\mbox{\tt A},\mbox{\tt C},\mbox{\tt G},\mbox{\tt
T},\mbox{\tt N}\}$ be the \emph{alphabet} of DNA sequences with a predefined
lexicographical order $\mbox{\tt \$}<\mbox{\tt A}<\mbox{\tt C}<\mbox{\tt
G}<\mbox{\tt T}<\mbox{\tt N}$, where `{\tt N}' represents an ambiguous base and
`{\tt \$}' is a sentinel that marks the end of a string. An element in $\Sigma$
is called a \emph{symbol} and a sequence of symbols is called a \emph{string}.
Given a string $T$, let $|T|$ be the length of the string, $T[i]$,
$i=0,\ldots,|T|-1$, be the $i$-th symbol in the string, $T[i,j]$, $0\le i\le
j<|T|$, be a substring and $T_i=T[i,|T|-1]$ be a suffix of $T$. Following
the definition by~\citet{en:2009fk}, we define a string terminated with `{\tt
\$}' as a \emph{text}. A text may have multiple sentinals. In a text $T$, if
$T[i]=\$$ and $T[j]=\$$, we mandate $T[i]<T[j]$ if and only if $i<j$. Thus when
we compare two suffixes of $T$, we do not need to compare beyond a sentinel
because each sentinel has a different lexicographical rank.

For two strings $P$ and $W$, let $P\circ W$ be their string concatenation.
We may sometimes write $P\circ W$ as $PW$ if it is unambiguous in the context.
Given an ordered set of texts, we call their ordered string concatenation
as a \emph{collection}, which is also a text. For example, suppose we have
two reads. The first is {\tt ACG} and the second is {\tt GTG}. The collection
of the two reads is $T=\mbox{\tt ACG\$GTG\$}$. Suffix $T_2<T_6$ because the first
sentinal is lexicographically smaller than the second.

For convenience, we assign an integer from 0 to 5 to `{\tt \$}', `{\tt A}',
`{\tt C}', `{\tt G}', `{\tt T}' and `{\tt N}', respectively. We may use both
the integer and the letter representations throughout the article. In addition,
given a symbol $a$, we define $\overline{a}$ as the Watson-Crick complement of $a$.
We regard the complement of `{\tt \$}' and `{\tt N}' is identical to itself.

\subsubsection{FM-index}

The \emph{suffix array} $S$ of text $T$ is a permutation of integers between 0 and
$|T|-1$, where $S(i)$, $0\le i<|T|$, is the position of the $i$-th smallest
suffix of $T$. Given a string $P$, the \emph{suffix array interval}
$I^l(P),I^u(P)]$ of $P$ in $T$ is defined as
\begin{eqnarray*}
I^l(P)&=&\min\{k:\mbox{$P$ is the prefix of $T_{S(k)}$}\}\\
I^u(P) &=&\max\{k:\mbox{$P$ is the prefix of $T_{S(k)}$}\}
\end{eqnarray*}
For convenience, we also define $I^s(P)=I^u(P)-I^l(P)+1$ as the size of the interval.

The \emph{Burrows-Wheeler Transform}~\citep{Burrows:1994gd}, or \emph{BWT}, of $T$ is a permutation of
symbols in $T$. The BWT string $B$ is computed as $B[i]=T[S(i)-1]$ for
$S(i)>0$ and $B[i]=\mbox{\tt \$}$ otherwise. Given a text $T$, also define
the accumulative count array $C(a)$ as the number of symbols in $T$ that are
lexicographically smaller than $a$, and the occurrence array $O(a,i)$ as
the occurrence of symbols $a$ in $B[0,i]$.

\emph{FM-index}~\citep{DBLP:conf/focs/FerraginaM00} is a compressed representation of the BWT $B$, the occurrence
array $O(a,i)$ and the suffix array $S(i)$. The key property of FM-index is
\begin{eqnarray}\label{eq:back}
I^l(aP)&=&C(a)+O(a,I^l(P)-1)\\
I^u(aP)&=&C(a)+O(a,I^u(P))-1
\end{eqnarray}
and $I^l(aP)\le I^u(aP)$ if and only if $aP$ is a substring of $T$.
We note that these two equations are different from the ones in our previous
paper~\citep{Li:2009uq} in that $C(a)$ and $O(a,i)$ defined here include
the sentinels, but the two arrays in the previous paper exclude them.

Given a collection $T=Q_0 Q_1\ldots Q_{n-1}$, we can retrieve
sequence $Q_i$ in linear time with Algorithm~1~\citep{DBLP:conf/recomb/MakinenNSV09}.
The second return value is the rank of $Q_i$ which equals $|\{Q_j:Q_j<Q_i\}|$.

\begin{algorithm}[h]
\DontPrintSemicolon
\footnotesize
\KwIn{Sequence index $i\ge0$}
\KwOut{Sequence $P$ and $k$, the rank of $P$}
\BlankLine
\textbf{Function} {\sc GetSeq}$(i)$
\Begin {
	$k\gets i$\;
	$P\gets\epsilon$\;
	\While{\bf true} {
		$a\gets B[k]$\;
		$k\gets C(a)+O(a,k)-1$\;
		\If{$a=0$}{
			\Return $(P,k)$
		}
		$P\gets aP$
	}
}
\caption{Sequence retrieval}
\end{algorithm}

\subsection{FMD-index}

Given DNA texts $R_0,\ldots,R_{n-1}$, define $T=R_0\overline{R}_0
R_1\overline{R}_1\ldots R_{n-1}\overline{R}_{n-1}$ as the
\emph{bidirectional collection} of the texts. We call the FM-index of $T$ as
the \emph{FMD-index} of $R_0,\ldots,R_{n-1}$ and define the \emph{bi-interval} of a
string $P$ as $[I^l(P),I^l(\overline{P}),I^s(P)]$. We will show how to compute
the bi-interval of $aP$ and $Pa$ when we know the bi-interval of $P$.

We note that when we know the bi-interval of $P$, $I^l(aP)$ and $I^s(aP)$ can
be readily computed with Equation~\eqref{eq:back}.
$[I^l(\overline{aP}),I^u(\overline{aP})]$ is a sub-interval of
$[I^l(\overline{P}),I^u(\overline{P})]$ because $\overline{P}$ is a prefix
of $\overline{aP}=\overline{P}\circ\overline{a}$. Due to the innate symmetry
of $T$, $I^s(\overline{cP})=I^s(cP)$ for all $c\in\Sigma$ with
$\sum_{c}I^s(cP)=I^s(P)=I^s(\overline{P})$. We can compute $I^s(cP)$ for all $c\in\Sigma$
with Equation~\eqref{eq:back}, use these interval sizes to divide
$[I^l(\overline{P}),I^u(\overline{P})]$ and finally derive
$[I^l(\overline{aP}),I^u(\overline{aP})]$.  This completes the computation of
the bi-interval of $aP$ (Algorithm~2).  Furthermore, when we backward extend
$P$, we actually forward extend $\overline{P}$.  Conversely, backward extension
of $\overline{P}$ yields forward extension of $P$ (Algorithm~3).
An FMD-index is bidirectional.

In comparison to the bidirectional BWT~\citep{Lam:2009fk} which uses two FM-indices,
the FMD-index builds both forward and reverse strand DNA sequences in one index.
Although the FMD-index is not applicable to generic texts, it is conceptually
more consistent with double-strand DNA and improves the speed of exact matching
as we only need to search against one index. For example,
BWA-SW~\citep{Li:2010fk} gets a 80\% speedup when we adopt the FMD-index as the
data structure.

\begin{algorithm}[h]
\DontPrintSemicolon
\footnotesize
\KwIn{Bi-interval $[k,l,s]$ of string $W$ and a symbol $a$}
\KwOut{Bi-interval of string $aW$}
\BlankLine
\textbf{Function} {\sc BackwardExt}$([k,l,s],a)$
\Begin {
	\For{$b\gets0$ \KwTo $5$} {
		$k_b\gets C(b) + O(b,k-1)$
		$s_b\gets O(b,k+s-1) - O(b,k-1)$
	}
	$l_0\gets l$\;
	$l_4\gets l_0+s_0$\;
	\For{$b\gets3$ \KwTo $1$} {
		$l_b\gets l_{b+1}+s_{b+1}$
	}
	$l_5\gets l_1+s_1$\;
	\Return{$[k_a,l_a,s_a]$}
}
\caption{Backward extension}
\end{algorithm}

\begin{algorithm}[h]
\DontPrintSemicolon
\footnotesize
\KwIn{Bi-interval $[k,l,s]$ of string $W$ and a symbol $a$}
\KwOut{Bi-interval of string $Wa$}
\BlankLine
\textbf{Function} {\sc ForwardExt}$([k,l,s],a)$
\Begin {
	$[l',k',s']\gets${\sc BackwardExt}$([l,k,s],\overline{a})$\;
	\Return{$[k',l',s']$}
}
\caption{Forward extension}
\end{algorithm}

\subsection{Unitig construction}

\subsubsection{Labeling reads and overlaps}
Given a bidirectional collection $T=R_0\overline{R}_0\ldots
R_{n-1}\overline{R}_{n-1}$, fermi labels the $i$-th input read $R_i$ with an
\emph{ordered} integer pair $[k,l]$, where $k$ is the rank of $R_i$
and $l$ the rank of $\overline{R}_i$. The pair $[k,l]$ can be computed by
{\sc GetSeq}$(2i)$ and {\sc GetSeq}$(2i+1)$, respectively. We also call
$[k,l]$ as the \emph{bi-interval of read} $R_i$. Obviously, the bi-interval
of read $\overline{R}_i$ is $[l,k]$, with the two integer swapped.

For two reads labeled by $[k,l]$ and $[k',l']$, if the tail (3'-end) of read
$[k,l]$ overlaps the head (5'-end) of $[k',l']$, we use an \emph{unordered}
integer pair $\langle l,k'\rangle$ to label the overlap. Such is a tail-to-head
overlap. Similarly, we use $\langle l',k\rangle$ for a head-to-tail overlap,
$\langle l,l'\rangle$ for tail-to-tail and $\langle k,k'\rangle$ for a
head-to-head overlap. The four types of overlaps correspond to the four
types of bidirectional edges in the bidirectional overlap
graph~\citep{Myers:1995nx}.

\subsubsection{Finding irreducible overlaps}
Finding irreducible overlaps plays a central role in fermi as well as in SGA.
Given its importance, we present a restructured version of this algorithm
(SD10;~\citealt{Simpson:2010uq}) using our notations (Algorithm~4).

In Algorithm~4, line 1 computes the bi-interval of a single symbol. The loop at
line 2 uses backward extensions to find all the reads overlapping with the
input string $P$. The loop at line 3 uses forward extensions base by base to exclude
reducible overlaps found at the previous step. $W$ is this loop keeps
the common substring of reads overlapping $P$ extended from the 3'-end of $P$. If 
in an iteration we find the sentinel of a read $R$ (line 5), then all the reads sharing
the same $W$ with $R$ must overlap with both $R$ and $P$ and therefore
their overlaps with $P$ are reducible. In this case, no further forward
extensions are necessary (line 4 and 6).

Similar to the original algorithm, Algorithm 4 requires that there are no contained reads.
Fermi actually implements a modified version that detects reads containment on the fly,
but we think the algorithm is a little overcomplicated. It is probably easier to
filter contained reads first and then run Algorithm 4, as SGA does.

\begin{algorithm}[h]
\DontPrintSemicolon
\footnotesize
\KwIn{Read $P$ and the minimum overlap length $x$}
\KwOut{Set of bi-intervals of reads having irreducible overlaps with the 3'-end of $P$}
\BlankLine
\textbf{Function} {\sc IrrOverlap}$(P,x)$
\Begin {
	\emph{Initialize ${\sf Curr}$ and ${\sf Prev}$ as empty arrays}\;
	$a\gets P[|P|-1]$\;
	\nl$[k,l,s]\gets [C(a), C(\overline{a}), C(a+1) - C(a)]$\;
	\nl\For{$i\gets |P|-2$ \KwTo $0$} {
		\If{$|P|-i-1\ge x$} {
			$[k',l',s']\gets${\sc BackwardExt$([k,l,s],0)$}\;
			\If{$s'\not=0$} {
				\emph{Append $([k',l',s'],\epsilon)$ to ${\sf Curr}$}\;
			}
		}
		$[k,l,s]\gets${\sc BackwardExt$([k,l,s],P[i])$}\;
	}
	\emph{Reverse array ${\sf Curr}$, and swap ${\sf Curr}$ and ${\sf Prev}$}\;
	${\sf Finished}=\emptyset$\;
	$\mathcal{I}=\emptyset$\;
	\nl\While{${\sf Prev}$ is not empty} {
		\emph{Reset ${\sf Curr}$ to empty}\;
		\For{$([k,l,s],W)$ {\bf in} ${\sf Prev}$} {
			\If{$W\in{\sf Finished}$} {
				\nl{\bf continue}\;
			}
			$[k',l',s']\gets${\sc ForwardExt}$([k,l,s],0)$\;
			\nl\If{$s'\not=0$} {
				${\sf Finished}\gets{\sf Finished}\cup \{W\}$\;
				$\mathcal{I}\gets\mathcal{I}\cup\{[k',l',s']\}$\;
				\nl{\bf continue}\;
			}
			\For{$a\gets1$ \KwTo $5$} {
				$[k',l',s']\gets${\sc ForwardExt}$([k,l,s],a)$\;
				\If{$s'\not=0$ {\bf and} $[k',l',s']$ is not in ${\sf Curr}$} {
					\emph{Append $([k',l',s'],Wa)$ to ${\sf Curr}$}\;
				}
			}
		}
		\emph{Swap ${\sf Curr}$ and ${\sf Prev}$}
	}
	\Return{${\sf IrrOvlp}$}
}
\caption{Finding irreducible overlaps (SD10)}
\end{algorithm}

\subsubsection{Unitig construction}

Unitig construction is a process of unambiguous merge of overlapped reads.
If $[k,l]$ and $[k',l']$ have an irreducible overlap $\langle l,k'\rangle$ and
can be unambiguously merged, we label the merged sequence with $[k,l']$; the
similar can be applied to other three types of overlaps.  With this simple
labeling procedure, we are able to fully keep track of the graph topology
during the unitig construction and without staging the graph in RAM.  This
procedure can also be easily multi-threaded.

\subsection{Finding the supermaximal exact matches}

An FMD-index can be used to find \emph{supermaximal exact matches} (SMEMs)
between a reference and a query sequence. Formally, a \emph{maximal exact
match} (MEM) is a an exact match that cannot be extended in either direction of
the match. An SMEM is a MEM that is not contained in other MEMs on the query
sequence. Fermi uses SMEMs to map reads back to the unitigs. 

Algorithm~5 describes the details. Basically, we use forward-backward extension
to extend an exact match and detect the boundary of a maximal match by tracking
the change of interval sizes. Fermi implements a variant of Algorithm~5. It
finds full-length read matches and can optionally exclude matches identical to
the query sequence.

\begin{algorithm}[h]
\DontPrintSemicolon
\footnotesize
\KwIn{String $P$ and start position $i_0$; $P[-1]=0$}
\KwOut{Set of bi-intervals of SMEMs overlapping $i_0$}
\BlankLine
\textbf{Function} {\sc SuperMEM1}$(P,x)$
\Begin {
	\emph{Initialize ${\sf Curr}$, ${\sf Prev}$ and ${\sf Match}$ as empty arrays}\;
	$[k,l,s]\gets [C(P[i_0]), C(\overline{P[i_0]}), C(P[i_0]+1) - C(P[i_0])]$\;
	\For{$i\gets i_0+1$ \KwTo $|P|$} {
		\If{$i=|P|$} {
			\emph{Append $[k,l,s]$ to ${\sf Curr}$}
		} \Else {
			$[k',l',s']\gets${\sc ForwardExt$([k,l,s],P[i])$}\;
			\If{$s'\not=s$} {
				\emph{Append $[k,l,s]$ to ${\sf Curr}$}
			}
			\If{$s'=0$} {
				{\bf break}\;
			}
			$[k,l,s]\gets[k',l',s']$
		}
	}
	\emph{Swap array ${\sf Curr}$ and ${\sf Prev}$}\;
	$i'\gets |P|$\;
	\For{$i\gets i_0-1$ \KwTo $-1$} {
		\emph{Reset ${\sf Curr}$ to empty}\;
		$s''\gets -1$\;
		\For{$[k,l,s]$ {\bf in} ${\sf Prev}$}{
			$[k',l',s']\gets${\sc BackwardExt$([k,l,s],P[i])$}\;
			\If{$s' = 0$ {\bf or} $i = -1$} {
				\If{${\sf Curr}$ is empty {\bf and} $i+1<i'+1$} {
					$i'\gets i$\;
					\emph{Append $[k,l,s]$ to ${\sf Match}$}
				}
			}
			\If{$s'\not=0$ {\bf and} $s'\not=s''$} {
				$s''\gets s'$\;
				\emph{Append $[k,l,s]$ to ${\sf Curr}$}
			}
		}
		\If{${\sf Curr}$ is empty} {
			{\bf break}
		}
		\emph{Swap ${\sf Curr}$ and ${\sf Prev}$}\;
	}
	\Return ${\sf Match}$
}
\caption{Finding super-maximal exact matches}
\end{algorithm}

\subsection{Other implementation details}

\subsubsection{Constructing FM-index}
To compute suffix arrays for strings with multiple sentinels, we modified an
optimized implementation of the SA-IS algorithm~\citep{DBLP:journals/tc/NongZC11} by Yuta
Mori.  We used the established algorithm to merge BWTs of subsets of
reads~\citep{DBLP:journals/algorithmica/HonLSSY07,en:2009fk,DBLP:conf/latin/FerraginaGM10}.
The BWT string is run-length encoded with the length in the delta
encoding~\citep{Elias:1975ly}.

\subsubsection{Error correction}
Fermi corrects potential sequencing errors using an algorithm similar to
solving the spectrum alignment problem~\citep{Pevzner:2001vn}, correcting
bases in underrepresented $k$-mers. It also shares similarity to
HiTEC~\citep{Ilie:2011fk}. Nonetheless, the fermi's algorithm differs in that
it is quality aware and does not replies on a user defined threshold on the
$k$-mer occurrences.

Fermi corrects errors in two phases. In the first phase, it collect all 23-mer
occurring 3 or more times using a top-down traversal over the trie represented
by the FMD-index. For each such 23-mer, fermi counts the occurrences of
the next (i.e. the 24-th) base and stores the information in a hash table
with the 23-mer being the key. In the second phase, fermi processes each read
by using the 23-mer hash table to correct errors by minimizing a heuristic cost
function of base quality and the occurrences of the 24-th base. Roughly
speaking, fermi tries to correct a low-quality base if by looking up its 23-mer
prefix we know the base is different from an overwhelmingly frequent 24-th base.
This algorithm can be adapted to correct INDEL sequencing errors in principle,
but more works are needed to perform minimization efficiently.

\subsubsection{Simplifying complex bubbles}
A \emph{bubble} is a directed acyclic subgraph with a single source and a
single sink having at least two paths between the source and the sink. A
\emph{closed bubble} is a bubble with no incomming edges from or outgoing edges
to other parts of the entire graph, except at the source and the sink vertices.
A closed bubble is \emph{simple} if there are exactly two paths between the
source and the sink; otherwise it is \emph{complex}.  In de novo assembly, a
bubble is frequently caused by sequencing errors or heterozygotes. Most
short-read assemblers uses a modified Dijkstra's algorithm to pop bubbles
progressively. Such an algorithm works fine for haploid genomes, but it is not
straightforward to distinguish heterozygotes from errors when the bubble is
complex.

Fermi uses a different algorithm. It effectively performs topological sorting
from the end of a vertex while keeping track of the top two paths containing
most reads. A bubble is detected when every path ends at a single vertex. It
then drops vertices not on the top two paths and thus turns a complex bubble to
a simple one.

\subsubsection{Using the paired-end information}
Given paired-end reads with short-insert sizes, fermi maps reads back to the
unitigs with Algorithm~5. If two unitigs are linked by at least five read
pairs, fermi will locally assemble the ends of unitigs together with unpaired
reads pointing to the gap under a relax setting. Fermi tries to align the
ends of unitigs using the Smith-Waterman algorithm, which may reveal imperfect
overlaps caused by sequencing errors or heterozygotes. Fermi also uses
paired-end reads to break contigs at regions without bridging read pairs. This
helps to reduce misassemblies during the unitig construction.
\end{methods}

\section{RESULTS}

We evaluated fermi on 101bp paired-end reads from NA12878~\citep{Depristo:2011vn}.
The total coverage of the original data is about 70-fold, but we only used half
of them. We assembled the 35-fold reads with fermi on a machine with 12 CPUs
and 96GB memory in about 5 days. The peak memory usage is 92GB.

We obtained unitigs of N50 1,022bp, totaling 3.83Gb. After collapsing
most heterozygotes and closing gaps with paired-end reads, we got longer contigs
(Table~4). Unitigs are short and redundant mainly because they break at
heterozygotes.

For SNP and INDEL calling, we aligned unitigs to the reference genome using
BWA-SW~\citep{Li:2010fk} with command line options `-b9 -q16 -r1 -w500'. We
called SNPs with the SAMtools caller and INDELs by directly counting INDELs
from the pileup output. We did not run a standard INDEL caller as short-read
INDEL callers do not work well with long contig sequences.

\subsection{Performance on de novo assembly}
We obtained the ALLPATHS-LG NA12878 contigs from NCBI (AC:AEKP01000000), the
SGA and SOAPdenovo scaffolds from
\mbox{\href{http://bit.ly/jts12878}{http://bit.ly/jts12878}}~\citep{Simpson:2011ly}, and the HuRef
assembly~\citep{Levy:2007uq} for
the comparison to the traditional capillary assembly. For the SGA and
SOAPdenovo scaffolds, we split at any ambiguous bases to get contigs; for the
HuRef assembly, we split at contiguous `N' longer than 20bp.

\begin{table}[bt]
\processtable{Statistics on human whole-genome assemblies}
{\begin{tabular}{lrrrrr}
\toprule
& AllPaths-LG & fermi & SGA & SOAP & HuRef \\
\midrule
Aligned contig bases & 2.62G & 2.82G & 2.74G & 2.71G & 2.88G \\
Aligned N50          & 22.6k & 15.6k & 9.8k  & 7.0k  & 81.4k \\
Covered ref. bases   & 2.59G & 2.74G & 2.70G & 2.67G & 2.78G \\
\# Type-1 breaks & 13,738 & 5,704 & 6,049 & 4,962 & 16,318 \\
\# Type-2 breaks & 3,823 & 1,120 & 1,735 & 727 & 6,626 \\
\botrule
\end{tabular}}{Contigs over 150bp in length are aligned to the human reference
genome GRCh37 with BWA-SW using option `-b33 -q50 -r17'. A type-1 break point
is detected if a contig is split in alignment and mapped to two distict locations, and at
each location the alignment is longer than 500bp and the mapping quality is no
less than 10. Type-2 break points exclude type-1 break points which can be patched
with gaps no longer than 500bp.} \end{table}

From Table~2, we can see that the HuRef assembly has much better
contiguity than short-read assemblies. It contains more
alignment break points when aligned against the GRCh37, but these are not
necessarily all misassemblies. ALLPATHS-LG has longer N50 than fermi, SGA and
SOAPdenovo partly because it uses 100-fold data and reads from jumping
libraries. However, the ALLPATHS-LG assembly covers fewer reference bases and
yields over twice as many as break points. Between fermi, SGA and SOAPdenovo
which are applied on the same data set, fermi has longer N50 with similar
number of alignment break points.  Overall, fermi achieves comparable assembly
quality to other assemblers.

\subsection{Performance on SNP and INDEL calling}

One of the key motivations of fermi is to explore the power of de novo assembly
in calling short variants. We collected several SNP and INDEL call sets (Table~3)
and compared the performance of fermi (Table~4 and 5).

\begin{table}[t]
\processtable{Evaluated SNP and INDEL call sets}
{\begin{tabular}{lllll}
\toprule
Label & Data & Assembler & Mapper & Caller \\
\midrule
AC & 96X Illumina PE$^1$ & AllPaths-LG & BWA-SW$^2$ & SAMtools$^3$ \\
BS & 70X Illumina PE & & BWA$^4$ & SAMtools \\
CG & Complete Genom.& & cgatools2$^5$ & cgatools2 \\
CV & 26X Illumina SE$^6$ & Cortex & & Cortex-var \\
FC & 35X Illumina SE$^6$ & fermi & BWA-SW$^2$ & SAMtools$^3$ \\
MD & 60X multiple        & & MAQ & 1000g pilot$^7$ \\
MI & Capillary reads$^8$     & &     & \\
SS & 35X Illumina SE$^6$ & & BWA-SW & SAMtools \\
\botrule
\end{tabular}}{
$^1$ AS uses reads from Illumina jumping and fosmid libraries\\
$^2$ BWA-SW is invoked with `bwa bwasw -b9 -q16 -r1 -w500'\\
$^3$ INDELs are called from pileup without using the SAMtools caller\\
$^4$ Realigned by GATK~\citep{Depristo:2011vn} also around known INDELs\\
$^5$ By Complete Genomics~\citep{Drmanac:2010ly}; only `VQHIGH' calls retained\\
$^6$ CV, FC and SS do not use the pairing information in calling\\
$^7$ 1000 Genomes Project pilot calls; generated from Dindel and multiple SNP callers\\
$^8$ INDEL calls by \citet{Mills:2011fk}\\
}
\end{table}

\begin{table}[b]
\processtable{Statistics of SNP call sets}
{\begin{tabular}{p{2.3cm}rrrrrrr}
\toprule
& FC & CV & SS & BS & CG & MD \\
\midrule
\#SNPs (M) & 3.37 & 2.20 & 3.24 & 3.50 & 3.34 & 2.69\\
\#hets (M) & 1.97 & 1.04 & 1.94 & 2.11 & 2.04 & 1.65\\
ts/tv      & 2.04 & 2.03 & 2.08 & 2.11 & 2.12 & 2.06\\
DN50 (bp)  & 3,593& 6,662& 3,523& 3,392& 3,447&3,992\\
DN2/DN50   & 22.3 & 20.8 & 23.4 & 22.7 & 22.3 & 22.9\\
\botrule
\end{tabular}}{Ts/tv is the transition-to-transversion ratio of SNPs. DN50 is
calculated as follows. The reference genome is masked according to the
align-ability mask~(\href{http://bit.ly/snpable}{http://bit.ly/snpable}) and segmented into intervals at
heterozygous SNPs. DN50 is computed such as 50\% of unique positions in the
genome are in intervals longer than DN50. DN2 is calculated similarly and
D2/DN50 is the ratio of DN2 and DN50. DN50 measures the sensitivity; the
smaller the better.  DN2/DN50 measures the precision of heterozygous SNPs; the
higher the better.}

\end{table}

For SNP calling (Table~4), fermi misses 3\% of SNPs called in SS,
but finds more additional ones. Manual examination reveals that
the additional calls are mainly caused by two factors. Firstly, in the
single-end mode, BWA-SW is very conservative. It may consistently give a
correct alignment a low mapping quality which are all downweighted by samtools.
Fermi is able to assemble such reads into
longer sequences which increase the power of BWA-SW. Secondly, in the fermi
alignment, some regions may be mapped with a high mismatching rate.  These may
be due to small-scale misassemblies in fermi unitigs or in the reference
assembly, or copy-number variations. It is possible that these clustered
SNPs contain more errors. Such errors may lead to reduced ts/tv, but tend
not to break long homozygous blocks due to very recent coalescences. That
is why FC has a high DN2/DN50 ratio, which measures how often false heterozygotes
arise from a long homozygous block.

Table~5 shows the comparison between different INDEL call sets. We excluded
INDELs around long homopolymer runs in all call sets because INDEL sequencing
errors tend to occur around long homopolymer runs and their error profile is still unclear
(the 1000 Genomes Project Analysis group, personal communication). In addition,
we have excluded the SS INDEL call set which is nearly contained in BS due to
the use of the same INDEL caller.

For the call sets in Table~5, MD and CG are relatively small due to the use of
very short reads. CV uses 26X 100bp reads. It is a small call set due to the
high false negative rate of the calling method~\citep{Iqbal:2012ys}. The fermi
call set FC is slightly smaller than BS, but it has larger overlap with other
call sets than BS, and more FC calls are confirmed by others. One explanation
to the lower overlapping ratio between BS and ALL is that BS is the only call
set that uses 101bp paired-end information, which gives it higher power for
INDELs not detectable with single-end or very short reads.  Nonetheless, purely
based on Table~5, fermi appears to have higher overall accuracy.


Even with all short-read call sets combined, as many as 14\% of double-hit
INDELs called by~\citet{Mills:2011fk} are missed. We manually checked 30
missing INDELs in an alignment viewer. For half of the cases, the short-read
alignment and fermi alignment strongly suggest no variations, and for all these
cases, the HuRef sequences are identical to GRCh37. In addition, there are a few
cases called from regions under clear copy-number changes. In all, we believe
INDELs called by~\citet{Mills:2011fk} only may have high error rate. With
short reads, we can recover most of short INDELs found by capillary sequencing.



\begin{table}[tb]
\processtable{Fraction of INDELs found in other call sets\label{tab:indel}}
{
\begin{tabular}{lrrrrrrr}
\toprule
   &     MD &     CG &     BS &     CV &     FC &     MI &    ALL \\
\midrule
MD & 240424 &  0.819 &  0.937 &  0.678 &  0.947 &  0.054 &  0.977 \\
CG &  0.752 & 264696 &  0.915 &  0.629 &  0.924 &  0.052 &  0.965 \\
BS &  0.564 &  0.597 & 404646 &  0.498 &  0.844 &  0.044 &  0.906 \\
CV &  0.708 &  0.726 &  0.882 & 251769 &  0.902 &  0.052 &  0.923 \\
FC &  0.588 &  0.624 &  0.873 &  0.522 & 393841 &  0.045 &  0.952 \\
MI &  0.593 &  0.618 &  0.790 &  0.527 &  0.804 &  23216 &  0.864 \\
\botrule
\end{tabular}}
{INDELs that start within a homopolymer run longer than 6bp are excluded in all
call sets. An INDEL in call set $R$ (indexed by row) is said to be \emph{found}
in call set $C$ (indexed by column) if there exists an INDEL in $C$ such that
the left-aligned starting positions of the two INDELs are within 20bp from each
other.  An INDEL in $R$ is considered to be found in `ALL' if it is found in
one of the other INDEL sets in the table, plus the AC call set.  In the table,
a number on the diagonal equals $|R|$, the number of INDEL calls in the call
set. The fraction equals $|\{g\in R:g \mbox{ is found in } C\}|/|R|$.}
\end{table}

\vspace*{-1em}

\section{Discussions}
In this article, we derived FMD-index by storing both forward and reverse
complement DNA sequences in FM-index. This simple modification enables faster
forward-backward search than bi-directional BWT~\citep{Lam:2009fk} and makes
FMD-index a more natural representation of DNA sequences. Based on FMD-index,
we developed a new de novo assembler, fermi, which achieves similar quality
to other mainstream assemblers.

We demonstrated that it is possible to call SNPs and short INDELs by aligning
assembled unitigs to the reference genome. This approach has similar SNP
accuracy to the standard mapping based SNP calling and arguably outperforms
the existing methods on INDEL calling in terms of both sensitivity and
precision. Assembly based variant calling is a practical and beneficial
complement to mapping based calling.

In the course of evaluating INDEL accuracy, we found that outside long
homopolymer regions, INDEL call sets do not often contain false positives, but
they may have high false negative rate, which leads to the apparent small
overlap between call sets~\citep{Lam:2012fk}.

As a theoretical remark, we note that with read counts kept, unitigs are a
\emph{lossless} but \emph{reduced} representation of sequence reads. They are
`reduced' in that individual reads are lost; they are `lossless' in that all
the information in reads, such as small variants, copy numbers and structural
changes are fully preserved in unitigs, as long as they are constructed
correctly. For single-end reads, it is theoretically possible to `compress'
reads to unitigs, which are largely non-redundant and much smaller in size.
Accurately and efficiently constructing unitigs might provide an interesting
alternative to data storage and downstream analyses in future, though practical
challenges, such as the high computational cost and the lack of accuracy of
unitigs, remain at present.

\vspace*{-1em}

\section*{ACKNOWLEDGEMENTS}
We are grateful to David Reich and Evan Eichler for providing additional data
for testing fermi, to Peter Sudmant for offering computing resources and
to David Altshuler and the Genome Sequence Analyis group for the helpful
discussions. We also thank Richard Durbin and Jared Simpson for their
comments on the initial draft of the manuscript.

\paragraph{Funding\textcolon} NIH 1U01HG005208-01.
\bibliography{fermi}

\end{document}